\documentstyle{article}
\begin{document}
\title{Statistics of conductance oscillations of a quantum dot in the 
Coulomb-blockade regime}
\author{F. Simmel, T. Heinzel$^*$, and D. A. Wharam\\Sektion Physik der 
Ludwig-Maximilians-Universit\"{a}t,\\
Geschwister-Scholl-Platz 1, D-80539 M\"{u}nchen}

\date{\today}
\maketitle
\begin{abstract}
The fluctuations and the distribution of the conductance peak spacings 
of a quantum dot in the Coulomb-blockade regime are 
studied and compared with the predictions of random matrix theory (RMT).
The experimental data were obtained in transport measurements performed 
on a semiconductor quantum dot fabricated in a GaAs-AlGaAs heterostructure.
It is found that the fluctuations in the peak spacings are considerably larger 
than the mean level spacing in the quantum dot. 
The distribution of the spacings appears Gaussian both for zero and for non-zero 
magnetic field and deviates strongly from the RMT-predictions.  
\end{abstract}
PACS numbers: 73.20.Dx,73.23.Hk,05.45.+b\\ \\
Advanced nanofabrication techniques have made it possible to confine
small numbers of electrons electrostatically within the two-dimensional 
electron gas (2DEG) of a semiconductor heterostructure 
\cite{houten91,meirav95}. 
Both the electric charge and energy of such ``quantum dots'' are 
quantised and hence such structures are sometimes referred to as
``artificial atoms'' \cite{kastner93,ashoori96}. 
In transport measurements the charging of these electron islands with
single electrons leads to the observation of periodic  
conductance oscillations in the Coulomb-blockade regime \cite{houten91}.
These reflect the electrostatic coupling of the quantum dot to its environment
and, additionally, they contain information about the eigenenergies and 
eigenfunctions of the electrons in the dot.
Due to irregularities in the electrostatic confinement potential and 
electron-electron interactions, the corresponding classical motion of the 
electrons in the quantum dot can be expected to be chaotic (nonintegrable)
\cite{jalabert92,stone93,bruus94}.
Consequently, recent experiments have considered the peak height 
distribution \cite{chang95,folk96}, parametric  conductance 
correlations \cite{folk96} and level statistics \cite{sivan96}
of a quantum dot in the Coulomb-blockade regime to test the concepts developed 
for the quantum mechanical description of classically chaotic systems 
(``quantum chaos'' \cite{gutzwiller90,haake90}). In particular random 
matrix theory (RMT) \cite{mehta91} has proven to be 
a very successful description
of the statistical properties of spectra of many irregular systems.
Therefore, it is a very interesting question, how RMT applies to the transport 
properties of quantum dots. In this paper we investigate the fluctuations 
of the peak spacings of the conductance peaks of a quantum dot 
obtained in transport measurements with and without applied magnetic field.
The spacing distributions are calculated and compared with the predictions 
of RMT.\\ 
For sufficiently low temperatures $T$ and small dot capacitances $C$, quantum 
dots isolated from the reservoirs of the 2DEG via 
tunnel barriers can exhibit Coulomb-blockade phenomena. 
When $e^2/C \gg k_BT$, transport through the quantum dot is blocked.
A finite conductance only occurs when the total energy of the quantum 
dot with $N$ electrons is degenerate with the energy of the dot occupied 
by $N+1$ electrons. This is the case when 
\begin{equation}
F(N+1)-F(N)=\mu,
\label{resonance}
\end{equation} 
where $F(N)$ denotes the free energy of the quantum dot with $N$ electrons 
and $\mu$ is the chemical potential of the leads. Then a single electron can 
tunnel from a reservoir into the dot \cite{houten91}. 
This can be achieved by tuning the dot's potential with a centre gate. 
A sweep in the centre gate voltage $V_g$ results in the well-known conductance 
oscillations in the Coulomb-blockade regime. 
From Eq.(\ref{resonance}) the difference $\Delta V_g$ between gate voltages
at which two adjacent peaks occur can be related to the thermodynamic 
quantity $\partial\mu / \partial N$, which has the meaning of 
an inverse compressibility \cite{sivan96}. 
Within the capacitive charging model \cite{houten91} 
the electrons are assumed to occupy single particle 
states of energies $\epsilon_i$ and the Coulomb interactions are described 
by a classical electrostatic term $U(N)$. The dot's energy is then 
$F(N) \approx \sum_i^N \epsilon_i +U(N)$  and the 
difference $\Delta V_g$ is given by 
\begin{equation}
e\alpha\Delta V_g^N=e^2/C+\Delta\epsilon_N.
\label{CB1}
\end{equation}
Here $e$ denotes the electronic charge, $C$ the total capacitance 
of the dot and $\Delta\epsilon_N=\epsilon_{N+1}-\epsilon_N$ the level 
spacing. The conversion factor $\alpha=C_g/C$, where 
$C_g$ is the dot-to-gate capacitance, translates between the energy 
and the voltage scale of the conductance oscillations. Thus, in 
principle, one should be able to extract the energy level
spacings $\Delta\epsilon_N$ from the so-called ``addition spectrum''
obtained in Coulomb-blockade measurements.\\
From the addition spectrum, one can calculate the nearest neighbour 
spacing (NNS) distribution $P(S)$, which can be compared to the 
predictions of RMT. $P(S)$ is the distribution of the spacings between
adjacent levels of an energy spectrum, where the spacings $S$ are 
normalised to a mean value of unity. The results for $P(S)$ within 
the framework of RMT are very well approximated by the Wigner 
surmise, which is \cite{haake90}
\begin{eqnarray}
P(S)&=&\frac{\pi}{2}S e^{-\frac{\pi}{4}S^2} \hspace{1.33cm} \mbox{(GOE)}
\label{GOE}\\
P(S)&=&\frac{32}{\pi^2}S^2 e^{-\frac{4}{\pi}S^2}\hspace{1cm} \mbox{(GUE)}
\label{GUE}
\end{eqnarray}
for time-reversal invariant systems and
for systems with broken time-reversal invariance, e.g. in the 
presence of a magnetic field. The first distribution corresponds to 
the energy spectrum of Hamiltonians drawn from the Gaussian 
orthogonal ensemble (GOE) of random matrices, while the second is
obtained for the Gaussian unitary ensemble (GUE). The fluctuations 
$\delta S=(\langle S^2\rangle - \langle S \rangle ^2)^{1/2}$ are thus 
expected to be $0.52 \langle S \rangle$ and 
$0.42 \langle S \rangle$ for GOE and GUE, respectively.\\ 
The quantum dot on which our measurements were performed was defined by
electron-beam lithography in the 2DEG of a GaAs-Al$_{0.32}$Ga$_{0.68}$As 
heterostructure. The mobility and the sheet density of the 2DEG are $120$m$^2/$Vs 
and $3.6\times10^{15}$m$^{-2}$, respectively. The application of 
negative gate voltages to the surface structure defines an island which 
is isolated from the left and right reservoirs via tunnel barriers (see 
inset of Fig.~\ref{CBosc}). The 
radius of the island is estimated to be $r\le 150$nm. The 
Coulomb-blockade measurements were performed in a dilution refrigerator 
with a base temperature of $25$mK. Electron transport through the dot was 
studied by applying a small bias voltage ($4.3\mu$V AC) between the left 
and right reservoirs and measuring the current using standard lock-in 
techniques (for further details see Ref.~\cite{heinzel94}).
From the onset of the conductance oscillations at $V_g=-200$mV
roughly $170$ peaks are observed (Fig.~\ref{CBosc}).
During the gate sweep from $V_g=-200$mV to 
$-1000$mV the number of electrons in the quantum dot thus varies from 
$N\approx 250$ to $N\approx 80$. 
In the following, the only relevant energy scale is 
the mean energy level spacing $\Delta$. It should be roughly $E_F/N$, 
where $E_F$ is the Fermi energy. From the sheet density one obtains 
$E_F \approx 12.9$meV and therefore $\Delta\approx 50\mu$eV. The thermal 
energy $k_BT$ is about one order of magnitude smaller.\\
To calculate the NNS distribution from the conductance oscillations, first the 
gate voltage differences $\Delta V_g$ between adjacent peaks are 
extracted from the data. The mean value of $\Delta V_g$ increases 
linearly with decreasing voltage  (see Fig.~\ref{spacing}(a)) 
reflecting an inverse linear change in the 
dot-to-gate capacitance \cite{chklovskii92}. 
Identifying $\langle \Delta V_g \rangle$ 
with the classical charging voltage $e/C_g$, from Eq.(\ref{CB1}) the 
energy spacings are obtained as
\begin{equation}
\Delta \epsilon=e\alpha(\Delta V_g-\langle \Delta V_g \rangle).
\label{CB2}
\end{equation}
The conversion factor $\alpha$ is also a function of the gate 
voltage. This can be considered by using the same linear fit as 
above, i.e. $\alpha=C_g/C=e(e+C_{rest}\cdot
\langle \Delta V_g \rangle)^{-1}$, where the capacitance $C_{rest}=C-C_g$ 
is assumed to be constant. However, the actual choice of $\alpha$
is not a crucial parameter in the calculation, as tests with different
constant values for $\alpha$ have shown. From Eq.(\ref{CB2}) 
the $\Delta\epsilon$ are obtained as fluctuations around a mean value 
of zero. To 
remove the unphysical negative values for $\Delta\epsilon$ the whole data 
are shifted by a constant value (cf. Fig.~\ref{spacing}(b)).
It turns out, that the fluctuations 
around the mean value are considerably larger than the mean level spacing 
$\Delta$ estimated above. This may be regarded as an indication that 
the calculated $\Delta \epsilon$ are not the ``real'' addition energies 
and that the influence of the electron-electron interactions both within 
the dot and its environment play a significant role \cite{sivan96}.\\
From the energy spacings one can construct an artificial one-particle 
energy spectrum via $E_i=\sum_{N=1}^i\Delta\epsilon_N$. To unfold the 
data to a mean level spacing of unity a polynomial fit is made to the 
spectral step function $N(E)=\sum_i \theta(E-E_i)$. The renormalized energies 
are obtained by the standard unfolding mapping $E_i \mapsto \langle 
N(E_i) \rangle$ \cite{gutzwiller90,haake90}. 
From these, the energy level spacings are calculated and 
can be directly compared to the predictions of RMT.\\
In Fig.~\ref{NNS1}(a) the resulting NNS distribution for zero magnetic field is 
displayed. It obviously does not agree with the Wigner surmise
(Eq.(\ref{GOE})). Instead, it is much better described by a Gaussian 
centered at $S=1$, as illustrated. 
\\ In the presence of a magnetic field, time-reversal invariance breaks down.
In this case the appropriate ensemble of random matrix theory to describe 
energy level fluctuations is the Gaussian unitary ensemble. However, as 
in the $B=0$T-case, the experimentally obtained spacing distributions
look Gaussian rather than GUE-like. In Fig.~\ref{NNS1}(b) the distributions 
for zero, for low ($B=0.1$T,$B=0.5$T) and high ($B=4$T) magnetic fields are 
displayed. The distributions are derived with typically 
$150-170$ data points. From a statistical point of view this number 
is rather small. Nonetheless, these are large numbers when compared to
previous Coulomb-blockade experiments \cite{sivan96}. It can be seen that 
the distributions narrow with increasing magnetic field as may be 
expected due to the Landau quantisation \cite{mceuen92,heinzel95}.\\ 
The largeness of the  fluctuations indicates that 
the capacitive term $e^2/C$ in Eq.(\ref{CB1}) undergoes even larger 
fluctuations than the energy levels themselves. 
Thus the $\Delta\epsilon$ obtained above 
do not display the energy level spectrum of the quantum dot.
However, this would not mean a failure of RMT, but an insufficiency of 
the capacitive charging model. Eq.(\ref{CB1}) obviously cannot be used to 
get access to the bare energy level spacings of the quantum dot, when a 
larger range of gate voltages is considered.\\
In a recent publication Sivan et al. \cite{sivan96} argued that 
electron-electron interactions in the dot were responsible for
the failure of RMT to describe the conductance peak spacing distribution.
Their experiments and calculations lead to a Gaussian $P(S)$ centered at $S=1$ 
which is similar to our results. In terms of the charging energy the 
fluctuations obtained in our experiment are $\delta(\Delta 
\epsilon) \approx 0.07 - 0.11 e^2/C$, which is slightly smaller than 
in the work by Sivan et al. Their numerical calculations suggest that 
fluctuations in the quantity $\Delta V/\langle \Delta V\rangle$ 
converge to a ``universal'' value between 0.1 and 0.2 for strong 
electronic interactions. Calculating this quantity from our data we arrive 
at 0.10, which is consistent with their finding. 
However, the influence of the capacitive coupling 
to the reservoirs has not been considered in their publication, which 
may also have a considerable influence on the fluctuation 
properties of the peak spacings.\\
Finally, it has to be considered that RMT was initially developed to 
handle the statistical properties of excitation spectra of complex systems.
The addition spectrum as obtained in Coulomb-blockade measurements, 
however, consists of the many-particle ground state energies of the 
quantum dot rather than excitation energies. The comparison with RMT has 
been made under the assumption that the addition spectrum be equivalent 
to a single particle spectrum. Indeed, the excitation spectrum of the 
model used in  \cite{sivan96}, obeys RMT. Recently, this could also be 
shown for the excitation spectrum of the two-dimensional Hubbard 
model \cite{bruus96}. But it is not clear whether the results of RMT can be 
applied to ground state energy statistics.
In this respect it is interesting to notice, that the peak 
height distribution for the conductance oscillations seems to be in
accordance with RMT \cite{chang95,folk96}, whereas
the parametric conductance correlations \cite{folk96,bruus96a} 
quantitatively do not agree with RMT.\\ 
In conclusion, we have investigated the statistics of conductance peak 
spacings obtained in Coulomb-blockade experiments with zero 
and non-zero magnetic field. In all cases the results do not agree with 
the predictions of random matrix theory. Instead, the nearest neighbour 
spacings appear to be Gaussian distributed around their mean value. 
It seems to be difficult to extract the bare energy levels when 
using a simple capacitive charging model. Therefore, our results include 
fluctuations in the electrostatic coupling with the environment which are 
larger than the fluctuations in the quantum dot's energy level spectrum itself.
Further theoretical and experimental work are required
to understand this central phenomenon in mesoscopic physics.\\ \\
We gratefully acknowledge stimulating discussions with  R. Berkovits, H. Bruus, 
W. H\"{a}usler, J. P. Kotthaus, C. Stafford and S. Ulloa and financial 
support from the Deutsche Forschungsgemeinschaft. One of us (F.S.) also 
acknowledges support from the Institute for Scientific Interchange 
Foundation.\\ \\
$^*$ present address: ETH Hoenggerberg HPF C13, CH-8093 Z\"{u}rich

\newpage 

\begin{figure}
\caption{Conductance oscillations of a quantum dot in the 
Coulomb-blockade regime at zero magnetic field
as a function of the centre-gate voltage $V_g$. Roughly $170$ peaks are 
observed between $V_g=-200$mV and $V_g=-1000$mV. The inset shows a 
schematic of the quantum dot. The shaded area denotes the 2DEG and the 
black areas indicate the gates with which the dot is defined. The lower 
middle gate is the centre-gate.}
\label{CBosc}
\end{figure}

\begin{figure}
\caption{(a) The peak spacings ($\Delta V_g$) extracted from Fig.~1 
(indicated by dots) 
and a linear fit to them. (b) Energy spacings ($\Delta \epsilon$) calculated 
from the peak spacings and shifted to positive values. The straight line 
indicates their mean value and the broken lines indicate
the largeness of the fluctuations expected from RMT.}
\label{spacing}
\end{figure}

\begin{figure}
\caption{(a) NNS histogram calculated from the energy spacings of Fig.~2(b) 
after unfolding them to a mean value of unity. The full line denotes the 
GOE-prediction of RMT for $P(S)$ and the broken line is a 
Gaussian fit centered at $S=1$.
(b) NNS distributions for different magnetic field strengths. The full 
and the broken lines denote the RMT-predictions for $P(S)$ obtained in 
the Gaussian orthogonal and unitary ensemble, respectively. For clarity, 
lines have been used to diplay the distributions instead of histograms.}
\label{NNS1}
\end{figure}

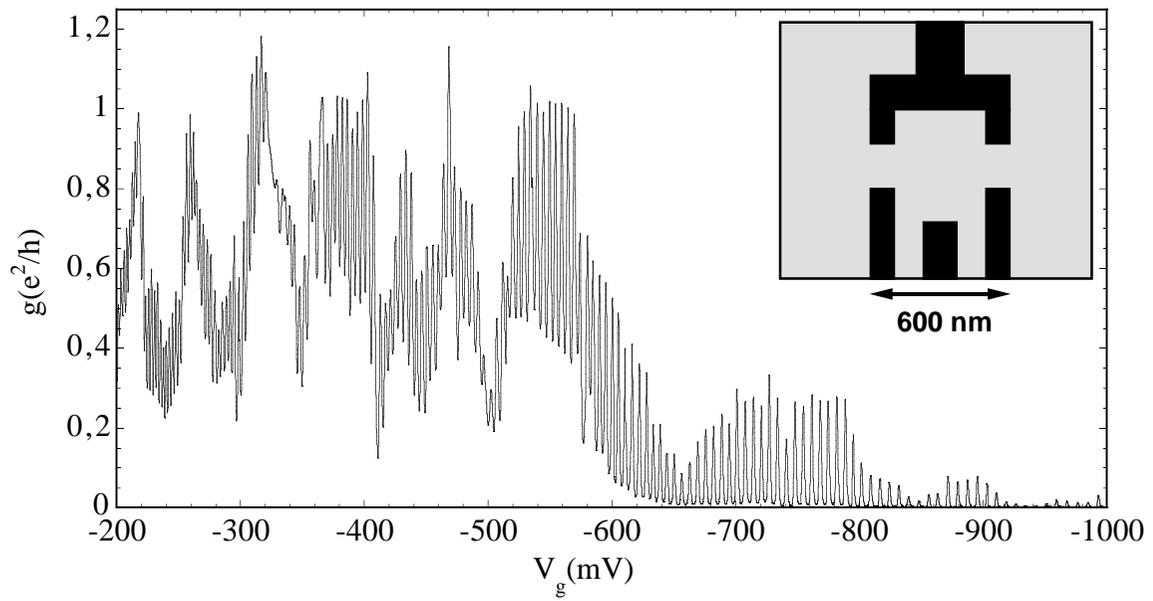
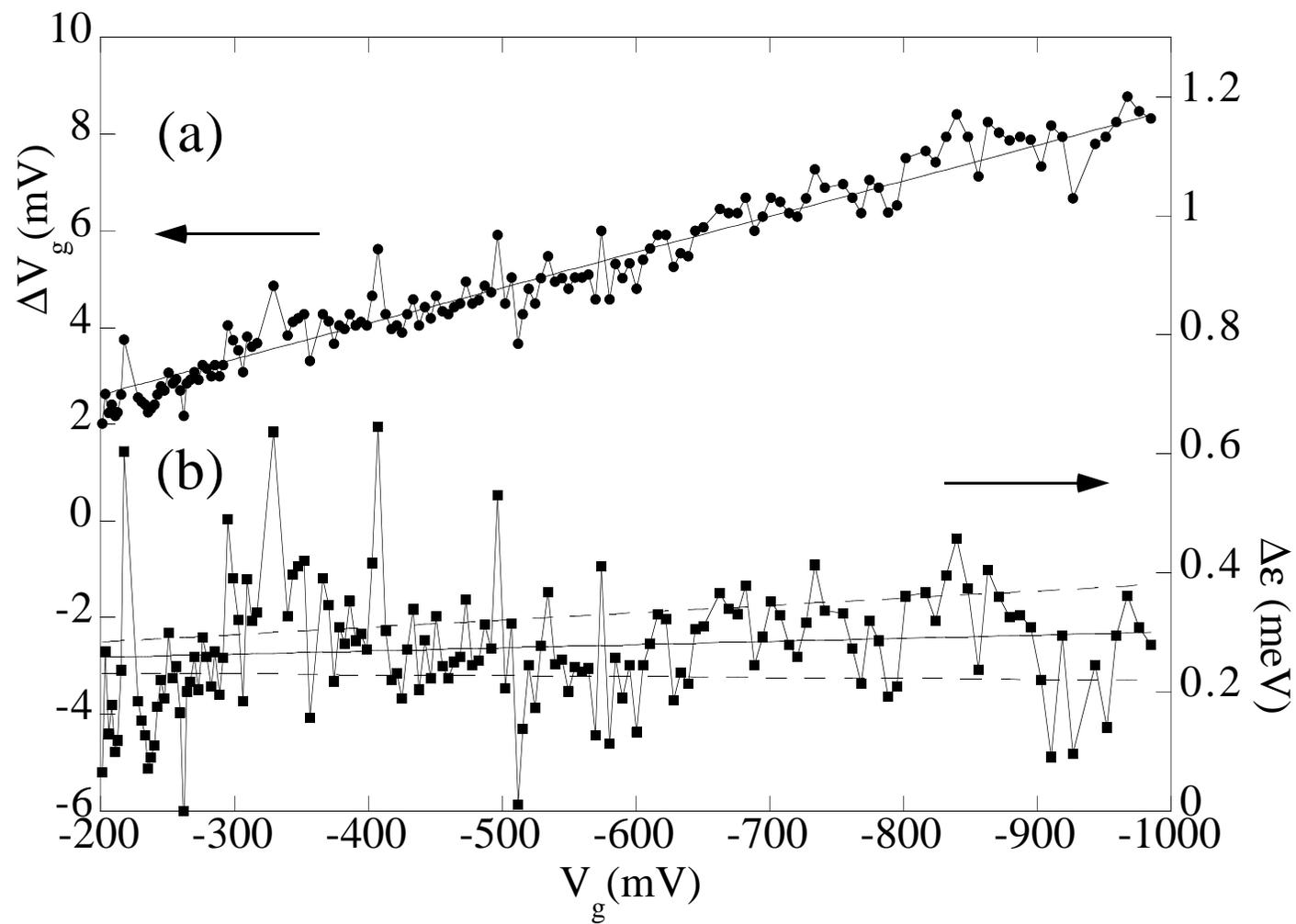
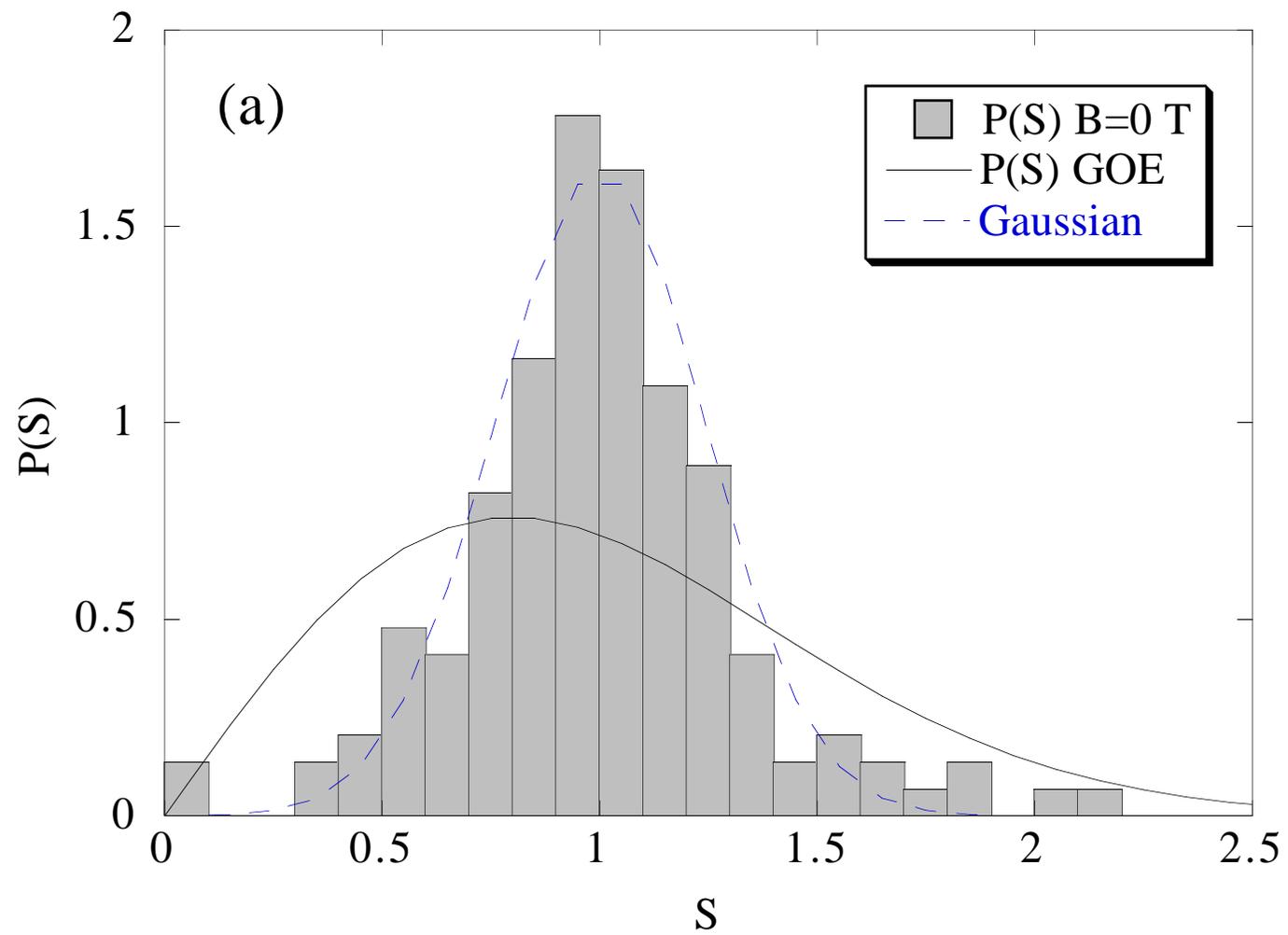
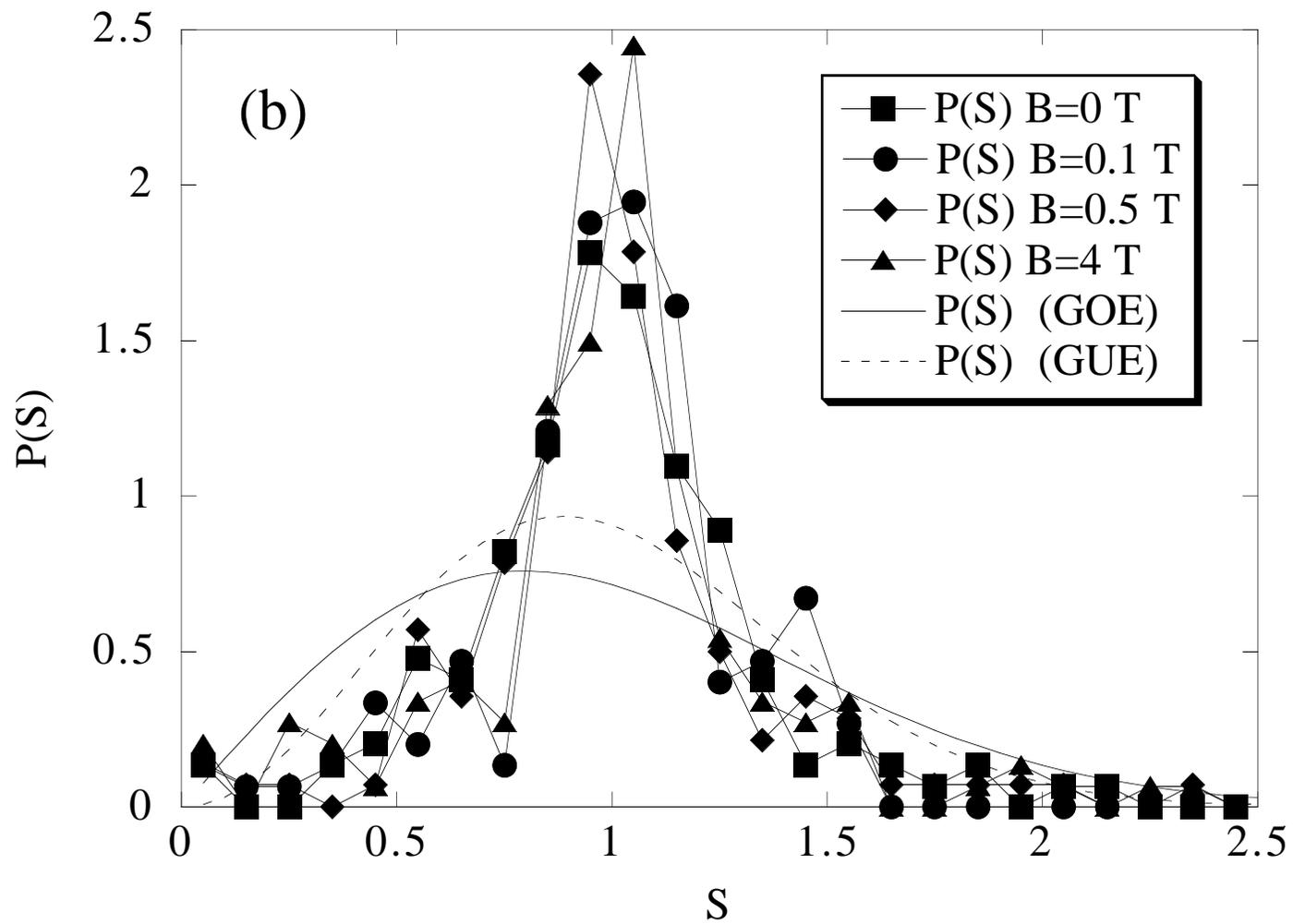
\end{document}